%Paper: hep-th/9508117
%From: Savdeep Sethi <sethi@string3.harvard.edu>
%Date: Wed, 23 Aug 95 23:04:30 -0400
%Date (revised): Wed, 23 Aug 95 23:19:41 -0400

%%
%% Revised draft of paper on monopole bound states in SUSY
%% Yang-Mills.
%%

\input harvmac.tex

\lref\rWSYMM{E. Witten,  Nucl. Phys. {\bf B202} (1982) 253.}
\lref\rSW{N. Seiberg and E. Witten, Nucl. Phys. {\bf B426} (1995)19
and Nucl.
Phys. {\bf B431} (1995) 484.}
\lref\rSMP{A. Sen, Phys. Lett. {\bf B329} (1994) 217, Int. J. Mod.
Phys. {\bf
A9} (1994) 3707.}
\lref\rNEWSUSY{S. Cecotti, P. Fendley, K. Intriligator and C. Vafa,
Nucl. Phys.
{\bf B386} (1992) 405. }
\lref\rCV{S. Cecotti and C. Vafa, Commun. Math. Phys. {\bf 158}
(1993) 569.}
\lref\rCasimirs{A. Perelomov and V. Popov, Sov. J. Nucl. Phys. {\bf
3} (???)
676.}
\lref\rMO{C. Montonen and D. Olive, Phys. Lett. {\bf B72} (1977)
117.}
\lref\rWB{J. Wess and J. Bagger, {\sl Supersymmetry and
Supergravity},
Princeton University Press (Princeton), 1992.}
\lref\rAH{M. Atiyah and N. Hitchin, {\sl Geometry and Dynamics of
Magnetic
Monopoles,} Princeton University Press (Princeton), 1988.}
\lref\rHowe{P. Howe, K. Stelle, and P. West, Phys. Lett. {\bf B124}
(1983) 55.}
\lref\rBlum{J. Gauntlett, Nucl. Phys. {\bf B411} (1994) 443 and J.
Blum, Phys.
Lett. {\bf B333} (1994) 92.}
\lref\rCallias{C. Callias, Commun. Math. Phys. {\bf 62} (1978), 213.}
\lref\rAG{L. Alvarez-Gaum\'{e}, Commun. Math. Phys. {\bf 90} (1983)
161 and J.
Phys. {\bf A16} (1983) 4177.}
\lref\rGSW{M. Green, J. Schwarz and E. Witten, {\sl Superstring
Theory, Vol
2.}, Cambridge University Press (Cambridge), 1987).}
\lref\rFW{D. Friedan and P. Windey, Nucl. Phys. {\bf B235} (1984)
395.}
\lref\rMS{N. S. Manton and B. J. Schroers, Ann. Phys. {\bf 225}
(1993) 290.}
\lref\rGM{G. W. Gibbons and N. S. Manton, Nucl. Phys. {\bf B274}
(1986) 183.}
\lref\rLM{H.B. Lawson and M-L. Michelsohn, {\sl Spin Geometry},
Princeton
University Press (Princeton),1989.}
\lref\rSF{M. Stern, J . Diff. Geom. {\bf 37} (1993) 467.}
\lref\rSS{M. Stern, Invent. Math. {\bf 115} (1994)  241.}
\lref\rBGV{N. Berline, E. Getzler, and M.Vergne, {\sl  Heat Kernels
and Dirac
Operators}, Springer-Verlag (New York), 1992.}
\lref\rJR{R. Jackiw and C. Rebbi, Phys. Rev.  {\bf D13}  (1976)
3398.}
\lref\rGMtwo{G. Gibbons and N. Manton,  ``The Moduli Space for Well
Separated
BPS Monopoles'', hep-th/9506052. }

%% Some definitions

\def\bps{Bogomol'nyi}
\def\YM{Yang-Mills}

\def\half{{1\over 2}}

\def\Mt{\widetilde{M}}

\def\frac#1#2{{#1\over #2}}

%
%       \eqn\label{a+b=c}	gives displayed equation, numbered
%				consecutively within sections.
%     \eqnn and \eqna define labels in advance (of eqalign?)
%
\def\eqnn#1{\xdef #1{(\secsym\the\meqno)}\writedef{#1\leftbracket#1}%
\global\advance\meqno by1\wrlabeL#1}
\def\eqna#1{\xdef #1##1{\hbox{$(\secsym\the\meqno##1)$}}
\writedef{#1\numbersign1\leftbracket#1{\numbersign1}}%
\global\advance\meqno by1\wrlabeL{#1$\{\}$}}
\def\eqn#1#2{\xdef
#1{(\secsym\the\meqno)}\writedef{#1\leftbracket#1}%
\global\advance\meqno by1$$#2\eqno#1\eqlabeL#1$$}

\Title{HUTP-95/A031, DUK-M-95-12}{\vbox{\centerline{Monopole and Dyon
Bound States in}
\vskip2pt\centerline{N=2 Supersymmetric Yang-Mills Theories}}}
\centerline{S. Sethi\footnote{$^1$}{Lyman Laboratory of Physics,
Harvard University, Cambridge, MA 02138, USA. Supported in part by
the Fannie
and John
Hertz Foundation (sethi@string.harvard.edu).}, M. Stern\footnote{$
^2$}{Mathematics Department, Duke University, Durham, NC 27708, USA.
Supported
in part by NSF Grant DMS 9505040.}  and E.
Zaslow\footnote{$ ^3$}{Mathematics Department, Harvard University,
Cambridge, MA 02138, USA. Supported in part by  DE-FG02-88ER-25065
(zaslow@math.harvard.edu).}  }

\vskip 0.5 in

We study the existence of monopole bound states saturating the BPS
bound in N=2
supersymmetric Yang-Mills theories. We describe how the existence of
such bound
states relates to the topology of index bundles over the moduli space
of BPS
solutions.  Using an $L^2$ index theorem, we prove the existence of
certain BPS
states predicted by Seiberg and Witten based on their study of the
vacuum
structure of N=2 Yang-Mills theories.

\Date{8/95}
\newsec{Introduction}

Strong-weak coupling duality, or S-duality, has been conjectured in
certain
supersymmetric field theories and string theories. The existence of
such a
symmetry was originally proposed by Montonen and Olive  for
Yang-Mills theories
\rMO. Among the testable predictions of S-duality is the existence of
certain
{\bps}-Prasad-Sommerfeld (BPS) bound states. For N=4 Yang-Mills, Sen
verified
the existence of such dyon bound states with magnetic charge two
\rSMP.  This
provided a strong dynamical test for the existence of  S-duality in
this
theory. For N=2 \YM\  theories, Seiberg and Witten have proposed a
generalized
S-duality involving the dependence of the theory on the Higgs field
expectation
value \rSW.  For the case of  N=2 \YM\ coupled to matter multiplets,
they
conjecture the existence of certain BPS bound states required by the
singularity structure of the vacuum manifold. We propose to verify at
least
some of their predictions.

Our approach to this problem involves quantizing the low-energy
dynamics of N=2
Yang-Mills coupled to matter. The existence of monopole bound states
then
reduces to a study of the spectrum of the Hamiltonian governing the
low-energy
dynamics. The existence of a bound state saturating the BPS bound is
then
equivalent to the existence of a zero mode for a twisted Dirac
operator on the
monopole moduli space. Since the moduli space is non-compact, we
employ an
$L^2$ index theorem to count the number of zero modes, and hence BPS
states at
magnetic charge two.

We briefly summarize our results.
N=2 Yang-Mills with $ N_f$ hypermultiplets has no BPS states at
magnetic charge
two for $ N_f <3$.  For $ N_f=3$, we find two bound states: one with
allowed
electric charges $ 4n+1$ and the other with allowed charges $ 4n+3,$
where $ n$
is an integer.\foot{The electric charge of an electron is normalized
to one as
in \rSW. } These states are singlets of the $ SO(6)$ flavor symmetry.
For the
candidate S-dual theory with $ N_f=4$, we find bound states
corresponding to
the $ SL(2,\bf{Z})$ partners of the fundamental electrons; however,
we also
find partners for the heavy gauge bosons, implying that they exist as
discrete
states at threshold in this theory.

In the following section, we derive the Lagrangian governing the
low-energy
dynamics of monopoles and dyons in supersymmetric Yang-Mills coupled
to matter.
Section three describes the moduli space of BPS solutions, and the
bundles of
interest to us. The index  computations needed to obtain the BPS
spectrum are
then presented. Section four provides a comparision with the states
predicted
by Seiberg and Witten. The final section is a summary and discussion.

\newsec{Supersymmetric Yang-Mills and the Collective Coordinate
Expansion}

\subsec{Pure N=2  \YM\ }

Collective coordinate expansions for N=2 and N=4 Yang-Mills around
BPS monopole
configurations have been described in \rBlum. We require the slightly
more
general case of N=2 coupled to general matter. Let us denote the $
SU(2)$
symmetry rotating the supersymmetry generators by $ SU(2)_I$. We
shall proceed
by first discussing the quantization of zero modes for pure N=2 \YM\
and then
coupling to matter. In N=1 superspace, the Lagrangian for N=2 \YM\
takes the
form

\eqn\LagYM{  L_{YM} = {1\over g^2} \int{d^4 \theta \Phi^{\dagger}
e^{2V} \Phi}
+ \left(  {1\over 4g^2}
      \int{d^2 \theta \Tr W^\alpha W_\alpha} + c.c. \right) , }
where $ W^\alpha$ is a vector multiplet, and $ \Phi$ a chiral
multiplet in the
adjoint representation of the gauge group. In terms of component
fields, the
Lagrangian contains a gauge field $ A^\mu$, Higgs field $ \phi$, and
an $
SU(2)_{I}$ doublet of
complex Weyl fermions $ \eta^{j}$.  Note that gauge indices are
suppressed for
most of this discussion. Let us take the gauge group to be $ SU(2)$
which
restricts any matter to the fundamental or adjoint representations.

In the Coulomb phase, the gauge group is broken from $ SU(2)$ to $
U(1)$. The
flat directions for the potential correspond to  $ \left[ \phi ,
\phi^{\dagger}
\right] = 0$. Let $ \phi$ have vacuum expectation value  $
\left\langle {\rm
Tr} \,\phi^2 \right\rangle = {1\over 2}v^2$ which we choose to be
large so
that a semi-classical analysis is applicable. In the Coulomb phase,
the theory
possesses fundamental particles and solitons with masses saturating
the BPS
bound. The BPS spectrum consists of dyons, W-bosons, and with the
inclusion of
matter, electrons. Some of these particles may exist at the quantum
level. As
explained in \rSW, the mass of such a state is determined by the
central
extension of the supersymmetry algebra, and is given by

\eqn\BPSmass{ M = \sqrt{2} | n_e v + n_m {4\pi i v\over g^2}  |}
in the semi-classical limit.  Here $ n_e , n_m$ are the electric and
magnetic
charges respectively. This formula remains true when matter is
coupled since we
shall assume all the electrons have vanishing bare masses.

Since our interest is in checking for the existence of BPS states, we
will
eventually consider only low-energy fluctuations around a BPS
solution. Such
fluctuations are tangential to the moduli space of BPS solutions with
charge
$k,$ and so our computations will reduce to non-relativistic quantum
mechanics
on the moduli space.

\subsec{BPS Field Configurations}

We shall follow the conventions of Wess and Bagger \rWB. The
supersymmetry
transformations for the action \LagYM\ take the form:

\eqn\susytransf{\eqalign{
\delta A_\mu  & = -i \bar{\eta_j} \bar{\sigma}_\mu \epsilon^j + i
\bar{\epsilon_j} \bar{\sigma}_\mu
\eta^j\cr
\delta \phi  \phantom{\,\,\,\,}&= \sqrt{2} \epsilon^j \eta_j\cr
\delta\eta^j \phantom{\,\,\,} &= \sigma^{\mu\nu} F_{\mu\nu}
\epsilon^j - i
\sqrt{2} \sigma^{\mu} D_{\mu} \phi \bar{\epsilon}^j \cr
}}
where we have set $ \left[ \phi , \phi^{\dagger} \right] $ to zero,
and where
the $ \epsilon^j_\alpha$ are anti-commuting parameters. We can
further choose $
\phi$ to be real and search for static field configurations which
preserve half
the supersymmetries. Such configurations satisfy the BPS bound. In
the gauge $
A_0=0$, with the magnetic field $ B_i = {1\over 2} \epsilon_{ijk}
F^{jk}$ , we
find the first order BPS equations

\eqn\BPSeqn{B_i = \pm \sqrt{2} D_i \phi.}
which also follow immediately from the classical energy for a
monopole
configuration
\eqn\energy{E = {1\over 2 g^2} \int{ (B_i \pm \sqrt{2} D_i \phi)^2}
\mp
{\sqrt{2} \over g^2} \int{\partial_i (B_i \phi)}.}
The second term is a topological invariant proportional to the
magnetic charge
and saturating the BPS bound \BPSmass.  For magnetic charge $k,$ a
field
configuration satisfying \BPSeqn\ depends on $4k$ parameters
$\varphi^a$. These
$4k$ moduli are coordinates for an interesting non-compact
hyperk\"{a}hler
manifold $M_k $ that will be discussed further in the following
section.
Associated to each modulus,  $\varphi^a$, are  zero modes $
(A^i_{oa},
\phi_{oa})$ in the expansion around the monopole configuration $
(A^i, \phi)$.
Such zero modes must be orthogonal to the directions generated by
infinitesimal
gauge transformations with compact support.  This condition
determines the
adjoint-valued parameters $ \omega_a$ defined by:

\eqn\omegaparam{\eqalign{
A^i_{0a} &=\partial_a A^i + D^i \omega_a\cr
\phi_{oa} &=  \left( \partial_a + \omega_a  \right) \phi .\cr
}}
The standard technique used to quantize these moduli is to permit the
$\varphi^a$ to vary with time while omitting the zero modes from a
perturbative
expansion. We will further truncate the mode expansion of  $ (A^i,
\phi)$ to
the moduli dependent classical solution.

We will expand the effective action to second order in the number of
time
derivatives and later to fourth order in the number of fermion
fields. To
ensure that the equations of motion are satisfied to first order, the
classical
configurations must be corrected by terms of order one; specifically,
the
constraint  $ A_0=0$ must be modified to satisfy the gauge-field
equations of
motion where now

\eqn\constraint{D_0 =\dot{ \varphi}^a\left( \partial_a + \omega_a
\right) .}
Taking into account such modifications, the bosonic piece of the
effective
action then describes a sigma model on the moduli space

\eqn\effaction{S_{eff} = \int{dt \, g_{ab} \dot{\varphi}^a\dot{
\varphi}^b},}
where
\eqn\metric{g_{ab} = {1\over g^2}\int{d^3 x \, ({1\over 2}A^i_{oa}
A^i_{ob} +
\phi_{oa} \phi_{ob})}.}
A more detailed discussion is given in \rBlum.

\subsec{Fermionic Zero Modes}

In the topologically non-trivial monopole background, the fermions
possess zero
modes. The inclusion of these zero modes in the effective action is
readily
accomplished by noting that the low-energy theory must be
supersymmetric since
the BPS configuration preserves a single supersymmetry. The Callias
index
theorem states that the Dirac operator for fermions in the adjoint
representation in a monopole background of charge $k$ has $2k$
zero modes \rCallias. We must therefore introduce $4k$ real Grassmann
collective coordinates $ \gamma^a$ in a mode expansion for the two
Majorana
fermions in the N=2 gauge multiplet. As in the case of the bosonic
moduli, we
allow the fermionic moduli to vary with time. By simply counting
degrees of
freedom, and using the pairing of bosonic and fermionic degrees of
freedom
required by \susytransf, we can easily determine the effective
action. The
action is most simply expressed in terms of  the superfields $ \Phi^a
=
\varphi^a + \theta \gamma^a$  where $ \theta$ is a Grassmann
coordinate \rFW:

\eqn\lowenergy{S_{eff} = \int{dt d\theta \, g_{ab} (\Phi)
\dot{\Phi}^a  D
\Phi^b}.}
The super-covariant derivative, $D,$ is given by $ D = - i {\partial
\over
\partial \theta} + \theta {\partial \over \partial t}$. This sigma
model
possesses more supersymmetries than are apparent from this superspace
formalism
since the moduli space is hyperk\"{a}hler, but we will not need to
make those
additional transformations explicit. Quantization of such a quantum
mechanical
model is a well-studied problem.  Supersymmetric ground states
correspond to
zero modes of the Dirac operator on the moduli space, or equivalently
to
anti-holomorphic closed forms on $ M_k$ since the space is Calabi-Yau
\rAG
\rGSW. We can now move easily to the case with matter multiplets.

\subsec{N=2 Yang-Mills Coupled to Matter}

The Lagrangian for an N=2 hypermultiplet contains two chiral N=1
superfields $
M$ and $ \widetilde{M}$ in conjugate representations of the gauge
group. With
bare masses set to zero, the matter Lagrangian is given by:

\eqn\matterlag{L_{M} = \int{d^4\theta \left(M^{\dagger} e^{2V} M +
\widetilde{M}^{\dagger} e^{-2V} \widetilde{M} \right)} +\left(
\sqrt{2}
\int{d^2\theta \widetilde{M}\Phi M } + c.c.  \right).}
The components of the hypermultiplet are an $ SU(2)_{I}$ doublet of
scalar
fields $ \left(m,  \widetilde{m}^{\dagger} \right)$ together with
complex Weyl
fermions $\left( \lambda , \widetilde{\lambda} \right)$ or
equivalently two
Majorana fermions. We shall discuss the case where $ M$ is in the
fundamental
representation of $ SU(2)$ with hermitian generators $ T^l$. In the
Coulomb
phase, the potential energy has no additional flat directions.
Further, there
are no zero modes for the scalars in the monopole background since
the operator
$$ - D^i D_i + (\sqrt{2} \phi^l T^l )^2 ,$$
is positive. However, the fermions do possess zero modes which form a
bundle
over the moduli space $ M_k$.  The bundle has dimension $ k$ and
transition
functions in $ O(k)$. For each Majorana fermion zero mode $
\lambda_{on}$, we
introduce a Grassmann collective coordinate $\psi^n,$ $n=1,...,k$. As
usual,
the $ \psi^n$ are dynamical and vary with time. The fermionic kinetic
term for
each Majorana fermions in the Lagrangian \matterlag\ provides a
natural bundle
metric $ h_{nm} (\varphi) = \int{ d^3 x \, \lambda_{0n} (\vec{x},
\varphi)
\gamma^0 \lambda_{0m} (\vec{x}, \varphi)}$ for the low-energy theory.
The other
ingredient needed to fully describe the effective action is a
connection for
the bundle. The connection also follows from the full action where we
recall
that $ A_0$ can now contain terms with two fermions as well as the
terms of
order one in time derivatives.  Let $ \widetilde{D}_0$ be the
restriction of $
D_0$ obtained by setting the fermions $ ( \gamma , \psi )$ to zero.
An
expression for the connection $ \Omega $ then follows from the term

\eqn\connection{\eqalign{
 \int{ d^3 x \, \lambda_{0n}  \gamma^0 \widetilde{D}_0 \lambda_{0m}
}&= \int{
\lambda_{on} \gamma^0  \dot{ \varphi}^a\left( \partial_a + \omega_a^l
T^l
\right) \lambda_{0m}}\cr
&= \dot{ \varphi}^a \Omega_{anm} ( \varphi) \cr
}}
where $ \omega_a$ satisfy \omegaparam.  The effective action can then
be
described using fermionic superfields $ \Psi^n = \psi^n + \theta F^n$
where the
second component is an auxillary field, and
\eqn\effaction{S_{eff} = \int{dt d\theta \, g_{ab} (\Phi)
\dot{\Phi^a}  D
\Phi^b + i h_{nm} ( \Phi ) \Psi^{n\alpha} (D \Psi^m_\alpha + D \Phi^a
\Omega_{ap}^m   \Psi^p_\alpha )}.}
The additional index $ \alpha$ affixed to the fermionic superfields
runs from $
1,\ldots , 2N_f$ since there are $ 2N_f$ Majorana fermions in the
fundamental
of an $ SO(2N_f)$ flavor symmetry described in \rSW. The symmetry
properties of
the bound states under this flavor group will be discussed further in
section
4. The supersymmetry generator for this theory corresponds to the
Dirac
operator coupled to the $ O(k)$ bundle connection. Supersymmetric
ground states
then correspond to normalizable zero modes of this twisted Dirac
operator.

The case that we will focus on for the remainder of the paper is
magnetic
charge $ k=2$. Since the connection has an abelian component in this
case,
states in the Hilbert space of this theory are labelled by their $
U(1)$
charge. Charge conjugation symmetry pairs a state with $ U(1)$ charge
$n$ to
one with charge -$n$. In the case of $ N_f$ hypermultiplets, the
spectrum of $
U(1)$ charges takes the form

\eqn\spectrum{ |n>, |n+1>, \ldots , |n+ 2 N_f>}
where $n$ is the $ U(1)$ charge of the Fock vacuum. Charge
conjugation  then
implies that $ n= - N_f$.  To summarize: the task of finding BPS
bound states
at magnetic charge two is equivalent to that of finding normalizable
zero modes
of the twisted Dirac operator acting on states with $ U(1)$ charge $
-N_f,
\ldots , N_f$.

\newsec{An Index Theorem for the Atiyah-Hitchin Manifold}

To count the number of normalizable zero modes of the twisted Dirac
operator,
we will employ index theory.  Clearly, we need to understand the
two-monopole
moduli space,  its metric, the bundle and its connection.
Fortunately, these
topics have been well-studied, and most of  the structures  we
require are
known.  We will benefit greatly from the work of \rAH,   \rGM\  and
\rMS\ in
this analysis.  However, the index computation is subtle since the
moduli space
is non-compact.  To count the number of $ L^2$ modes will require a
careful
analysis of boundary effects. First however, we must describe the
geometry of
the moduli space.

\subsec{Monopole Moduli Space}

As discussed in section 2.1, the BPS equations \BPSeqn\ for magnetic
charge k
have families of  solutions with $ 4k$ parameters.  For $k=1$, these
parameters
describe  translations and ``large gauge transformations" of the
standard BPS
monopole solution. The moduli space in this case is just ${\bf
R}^3\times S^1$.
 For general $k$ the space has the form

\eqn\modulispace{M_k = {\bf R^3}\times{S^1\times M_k^0 \over {\bf
Z}_k }, }
where $M_k^0$ is a $4k-4$ real dimensional hyperk\"{a}hler manifold
equipped
with a hyperk\"{a}hler metric \rAH.  It is often useful to consider
the $k-$fold cover $\widetilde{M}_k = {\bf R}^3\times S^1\times
M_k^0.$  The
metric on $\widetilde{M}_k$ is flat in the ${\bf R}^3\times S^1$
factors, so
we will focus our discussion on $M_k^0$ -- specifically, $M_2^0.$

Invariance of the BPS equations under the Euclidean group amounts to,
in part, an $SO(3)$ group of isometries on $M_2^0,$ the generic orbit
of which is three dimensional.  We identify $so(3)$ with the left
invariant
vector fields on $SO(3)$ in the usual manner. Let $\{X_1,X_2,X_3\}$
be a basis
for $so(3)$ satisfying
$$[X_1,X_2] = -X_3,  \;\; [X_3,X_1] = -X_2, \;\; [X_2,X_3] = -X_1,$$
and let   $\{\sigma_i\}$ denote the dual frame. They satisfy the
relation
$d\sigma_i = \half\epsilon_{ijk}\sigma_j\wedge\sigma_k.$

In this notation, the metric on $M_2^0$ is constrained to be of the
form
\eqn\modmet{ds^2 = f(r)^2dr^2 + a_1(r)^2\sigma_1^2
+a_2(r)^2\sigma_2^2
+a_3(r)^2\sigma_3^2 .}
 Where convenient, the $\sigma_i$ will be described by $SO(3)$ Euler
angles
$0\leq \theta < \pi, 0\leq\phi < 2\pi,0\leq\psi < 2\pi:$
$$\matrix{
\sigma_1 & =  & -\sin\psi d\theta + \cos\psi \sin\theta d\phi \cr
\sigma_2 & =  & \cos\psi d\theta + \sin\psi \sin\theta d\phi   \cr
\sigma_3 & =  & d\psi  + \cos\theta d\phi.\cr
}$$

   $M_2^0$ also has the identification
\eqn\identify{(r,\theta,\phi,\psi)\equiv (r,\pi -\theta,\phi + \pi,
-\psi).}
Anti-self-duality of the curvature -- following from
hyperk\"{a}hlerity --
tells us that
$$  {2a_2 a_3\over f}{ da_1\over dr} = (a_2-a_3)^2 -a_1^2,$$
as well as the equations obtained by cyclically permuting
$a_1,a_2,a_3.$

By redefining the radial coordinate if necessary, we can take
$f=-a_2/r,$ and
the range
of $r$ is then from $\pi$ to $\infty.$ For this choice the large $r$
dependence
of the metric is found to be
$$a_1  \approx a_2 \approx r\sqrt{1-{2\over r}}  $$
$$a_3 \approx -{2\over \sqrt{1-(2/r)} } $$
up to terms suppressed by $e^{-r}$ \rGM.
Near the bolt  coordinate singularity at $r=\pi$, we have
$$
\eqalign{ a_1 & \approx   2(r-\pi) \cr
a_2 & \approx   \half(r-\pi) +\pi  \cr
a_3 & \approx    \half(r-\pi) -\pi. \cr
} $$

As discussed in the previous section, the zero energy solutions to
the Dirac
equation form an $ O(k)$ bundle over the moduli space $ M_k$. The
bundle is
trivial over the $  {\bf R^3}$ factor in the decomposition of the
moduli space
given by \modulispace. Further, Hitchin has shown that the connection
on this
bundle has anti-self-dual curvature.\foot{As referred to in \rMS.}
Let us
first consider the case $ k=1$ where the index bundle is an $ O(1)$
bundle.
While obviously flat, the bundle is not trivial, and is once twisted
over the $
S^1$ factor.  We now focus our attention on the nontrivial case $
k=2$.  Let
${\rm Ind}_2$ denote the restriction of the bundle to $(S^1\times
M_2^0)/{\bf
Z}_2$. The obstruction to orienting the bundle arises from the
explicit
discrete $ {\bf Z}_2$ identification in \modulispace, which we call
$I_3.$  In
coordinates, $I_3$ is the identification

\eqn\ithree{ (\chi; r,\theta,\phi,\psi ) \equiv (\chi + \pi;
r,\theta,\phi,\psi
+\pi ),}
where $ 0 \leq \chi < 2\pi$ is a coordinate for $ S^1$. We can
therefore pull
back this bundle to  $S^1\times M_2^0$ and make a choice of
orientation giving
a $ U(1)$ bundle $\widetilde{\rm Ind}_2 $.

The anti-self-duality of the bundle curvature, and its invariance
under the $
SO(3)$ subgroup of the Euclidean group of symmetries of the BPS
equations, fix
the curvature to lie entirely in $M_2^0$ and have the form

\eqn\bundcurv{\Omega = d\alpha\wedge\sigma_1 +
\alpha\sigma_2\wedge\sigma_3.}
The function $\alpha (r)$  falls off as $e^{-r/2}$ as
$r\rightarrow\infty$, and
has a normalization $\alpha(\pi)=\pm\half$ \rMS. The sign ambiguity
depends on
the choice of orientation, and will play no role in the following
analysis.

\subsec{The Index Formula}

In this section, we consider the restriction of $\widetilde{\rm
Ind}_2$ to
$(p,M_2^0) \cong M_2^0,$
where $p$ is any point on $S^1.$\foot{Though $\widetilde{\rm Ind}_2$
wraps
nontrivially
around the $S^1,$ there is no ambiguity in defining $I,$ up to
isomorphism.}
$I$ inherits a connection from $\widetilde{\rm Ind}_2.$
We wish to compute the dimension of the kernel of the Dirac operator
$D$ on
spinors
with values in $I^n.$  First we need a vanishing theorem.

\vskip 0.1in
{\noindent \bf Proposition 3.2.1} {\it
Let $M$ be an infinite volume spin four-manifold with zero scalar
curvature.
Let
$E$ be a line bundle over $M$ whose curvature $r^E$ is an
anti-self-dual
two-form.  Then
the kernel of the Dirac operator $D_E^-$ on $S^{-}\otimes E$ is
zero.}

\noindent
Proof. Integrating by parts and using the Bochner-Lichnerowicz
formula (\rLM,
Theorem 8.17)
we have
$$\|D_E^-f\|^2 = \|\nabla f\|^2 + (R^Ef,f),$$
where $R^E = \sum_{i<j}e_ie_jr^E_{i,j},$ and $e_i$ denotes Clifford
multiplcation by the
$i^{th}$ vector in a frame.
In an oriented frame, we may write the projection onto $S^{-}\otimes
E$ as
 $(1 - e_0e_1e_2e_3)/2$.  The anti-self-duality of $r^E$ implies
$$\sum_{i<j}e_ie_jr^E_{i,j}(1 - e_0e_1e_2e_3)/2 = 0.$$
Hence,
$$\|D_E^-f\|^2 = \|\nabla f\|^2,$$
for $f$ a section of $S^{-}\otimes E$.  An element of the kernel of
$D_E^-$ is
therefore
covariantly constant. This implies its norm is covariantly constant.
On an
infinite volume
manifold the only covariantly constant square integrable section is
zero.
${\,\lower0.9pt\vbox{\hrule \hbox{\vrule height 0.2 cm \hskip 0.2 cm
\vrule
height 0.2 cm}\hrule}\,}$
\vskip 0.1in
{\noindent \bf Corollary 3.2.2} {\it
The $L^2$ index of the Dirac operator
$$D^+: L^2(M_2^0,S^+\otimes I^n)\rightarrow L^2(M_2^0,S^-\otimes
I^n)$$ is the
dimension
of the kernel of $D^+$.}
\medskip
Thus we are left with the computation of the $L^2$ index. We recall
the
modifications required to compute the index on a complete noncompact
manifold.
Let $e^{-sD^2}$ denote the heat operator associated to $D^2$.  As
$s\rightarrow\infty$,  $e^{-sD^2}$ converges weakly to projection,
$\Pi$, onto
the kernel of $D$. Hence, the integral of
$tre^{-sD^2}(x,x)$ over a compact subset $C$  converges to
$\int_Ctr\Pi(x,x)dx,$ and  $\int_C tr\tau e^{-sD^2}(x,x)dx
\rightarrow \int_C
tr\tau\Pi(x,x)dx,$
where $\tau $ denotes Clifford multiplication by the volume element.
$\tau = e_0e_1e_2e_3$ in the notation of the preceding proposition.
The $L^2$ index of $D$ is given by the integral
$${\rm Ind} (D) = \int_{M_2^0}tr\tau\Pi(x,x)dx =
\int_{M_2^0}lim_{s\rightarrow\infty}tr\tau e^{-sD^2}(x,x)dx.$$
On a compact manifold, $Y,$ with Dirac operator $D_Y$,
$\int_Y tr\tau e^{-sD_Y^2}(x,x) dx$
is independent of $s$; hence,  in the above discussion the
$s\rightarrow\infty$
 can be replaced by $s\rightarrow 0$, and one obtains
$${\rm Ind} (D_{Y}) = lim_{s\rightarrow 0}\int_{Y}tr\tau
e^{-sD_Y^2}(x,x)dx.$$
 In the latter limit, the traces are easily computed.
In general $\frac{d}{ds}tr\tau e^{-sD^2}(x,x)$ can be expressed as
the
divergence of some vector field $V(s)$,
with
$$\int_C divV(s)dx = \int_{\partial C}tr e_{0}\tau
De^{-sD^2}d\sigma.$$
Here $e_{0}$ is Clifford multiplication by the unit outward normal to
$\partial C$ and $d\sigma$ is the induced volume form on $\partial
C.$
(See \rCallias.)  This explains the $s$ independence of the trace in
the
compact case and yields the following expression for the index on
$M_2^0$
(see, e.g. \rSF, \rSS, and \rCallias):
$${\rm Ind}(D) =
lim_{L\rightarrow\infty}(lim_{s\rightarrow 0}\int_{r<L}\tr \tau
e^{-sD^2}dx
+ \int_{0}^{\infty}\int_{r = L} tre_0\tau De^{-sD^2}d\sigma ds).$$
The Atiyah-Singer index theorem gives

$$\eqalign{ & lim_{L\rightarrow\infty}(lim_{s\rightarrow
0}\int_{r<L}\tr \tau
e^{-sD^2}dx )
=  \cr &  {1\over 24\cdot  8\pi^2}\int_{M_2^0}\Tr(R\wedge R) +
{1\over
8\pi^2}\int_{M_2^0}\Tr(\Omega\wedge\Omega).}$$
We  write

 \eqn\index{ {\rm Ind} (D) =  {1\over 24\cdot
8\pi^2}\int_{M_2^0}\Tr(R\wedge R) +
{1\over 8\pi^2}\int_{M_2^0}\Tr(\Omega\wedge\Omega) + \delta_D,}
where
$$\delta_D = lim_{L\rightarrow\infty}\int_{0}^{\infty}\int_{r = L}
tre_0\tau
De^{-sD^2}d\sigma ds.$$

{}From the form of the metric \modmet, one computes the curvature in
the
orthonormal frame
$$\{\nu_0,\nu_1,\nu_2,\nu_3\} \equiv
\{f\,dr,a_1\sigma_1,a_2\sigma_2,a_3\sigma_3\}$$
to be
$$\matrix{
R_{10}  & = & R_{23} & = & {1\over a_1f} ({\partial \over \partial
r}{A}
)\nu_{0}\wedge \nu_{1} + {1\over a_2 a_3} (A+B+C-1-2BC) \nu_2\wedge
\nu_3,\cr
R_{20}  & = & R_{31} & = & {1\over a_2 f} ({\partial \over \partial
r}{B}
)\nu_{0}\wedge \nu_{2} + {1\over a_1 a_3} (A+B+C-1-2AC) \nu_3\wedge
\nu_1,\cr
R_{30}  & = & R_{12} & = &{1\over a_3f} ({\partial \over \partial
r}{C}
)\nu_{0}\wedge \nu_{3} + {1\over a_1 a_2} (A+B+C-1-2AB) \nu_1\wedge
{\nu_2},
\cr
}
$$
where
$$A = (a_2^2 + a_3^2 - a_1^2)/2a_2a_3,$$
$$B = (a_3^2 + a_1^2 - a_2^2)/2a_3a_1,$$
$$C = (a_1^2 + a_2^2 - a_3^2)/2a_1a_2.$$
Anti-self-duality holds for the choice of orientation
$\nu_0\wedge \nu_1\wedge \nu_2\wedge \nu_3$, and we see that
$$ \Tr (R\wedge R) = d\left[ \left( -4(A+B+C-1)^2 + 16ABC\right)
\sigma_1\wedge\sigma_2\wedge\sigma_3\right], $$
which by Stokes' theorem (note that the boundary orientation induced
here
is $-\nu_1\wedge \nu_2\wedge \nu_3$) and the asymptotic formulas for
$a_1,a_2,$
and $a_3$ gives
$${1\over 24\cdot 8\pi^2}\int_{M_2^0} \Tr(R\wedge R) = -{1\over 6}.$$
Note that we have divided by a factor of two due to the ${\bf Z}_2$
identification \identify.

The next term in the bulk contribution is also simple to compute,
since
we have, from \bundcurv,
$$\Omega\wedge\Omega = n^2 d(\alpha^2\wedge\sigma_1\wedge\sigma_2
\wedge\sigma_3)$$
for the bundle $I^n$.
Using $\alpha^2(\infty) = 0$ and $\alpha^2(\pi) = 1/4,$ we find
$${1\over 8\pi^2}\int_{M_2^0}\Tr(\Omega\wedge\Omega) = {n^2\over
8}.$$
Combining the terms, we arrive at
\eqn\definition{{\rm Ind}(D) = n^2/8 - 1/6 + \delta_D.}
We now turn to the calculation of the boundary contribution
$\delta_D$.

\subsec{An Equivalent Index Problem}
  For $K$ some large positive constant, equip $(K,\infty)\times
SO(3)$ with the
metric $f(r)^2dr^2 + \sum a_i^2(r)\sigma_i^2$.  This metric descends
to a
metric on $(K,\infty)\times SO(3)/{\bf Z}_2,$ where the ${\bf Z}_2$
action is
the one generated by \identify.  This is just the Atiyah-Hitchin
metric
described in \modmet\ near $\infty$.

In order to reduce the computation of the index of the Dirac operator
to
computations
essentially the same as those carried out in  \rSF,  we will first
make a
conformal change in the metric of the moduli space in a neighborhood
of
$\infty$. This change is not essential to computing the defect
$\delta_D $ but
simplifies the estimation of error terms involved in
constructing approximations to resolvents and heat operators.  By the
conformal
invariance of the Pontrjagin classes, we know that such a conformal
change
will not change the value of the bulk contribution.  Hence, in order
to justify
it,
we need only check that it preserves the index.  First let us specify
a new
metric which, for large $ r$, is given by

\eqn\newmetric{\eqalign{
 g &= {f(r)^2\over a_2^2} dr^2 +
\sum_1^3\frac{a_i^2}{a_2^2}\sigma_i^2 \cr
    &= \frac{dr^2}{r^2} + (1+A_1)\sigma_1^2 + \sigma_2^2 +
(\frac{4}{r^2(1-2/r)^2} + A_3)\sigma_3^2,\cr
}}
with $A_j$ doubly exponentially decreasing functions of $t$ (i.e.
$O(e^{kt}e^{-e^t})$).  Set $t = ln(r).$ The metric is then

$$dt^2 + (1+A_1)\sigma_1^2 + \sigma_2^2 +
(\frac{4}{e^{2t}(1-2e^{-t})^2} +
A_3)\sigma_3^2.$$
We fix an orthonormal frame $\{Y_i\}$ on the ${\bf Z}_2$ cover of our
space
with
$Y_0 = \frac{\partial}{\partial t},$ $Y_1 = (1+A_4)X_1$, $Y_2 = X_2,$
and $Y_3
= ( e^{t}/2 - 1 + A_5)X_3$, with $A_4,A_5$ doubly exponentially
decreasing. Of
course this frame does not  descend to one defined globally on the
${\bf Z}_2$
quotient, but that will not affect our computations.  Quantities
involving the
squares of these vector fields will descend to the quotient.

Neglecting doubly exponentially decreasing terms, we have the
commutation
relations:

$$
\eqalign{
[Y_1,Y_2] & \sim -\frac{2e^{-t}}{1-2e^{-t}}Y_3, \cr
[Y_3,Y_1] & \sim -e^{t}(1/2-e^{-t})Y_2, \cr
[Y_2,Y_3] & \sim -e^{t}(1/2-e^{-t})Y_1, \cr
[Y_0,Y_3] & \sim (1-2e^{-t})^{-1}Y_3.
} $$
{}From these relations, we can check modulo doubly exponentially
decreasing
terms that
$$
\eqalign{
(\nabla_{Y_3}Y_1,Y_2) & =  -(e^t/2 - 1 - (e^t/2 - 1)^{-1}/2), \cr
(\nabla_{Y_1}Y_2,Y_3) & = -(e^t - 2)^{-1}, \cr
(\nabla_{Y_2}Y_3,Y_1) & = (e^t - 2)^{-1}, \cr
(\nabla_{Y_3}Y_0,Y_3) & = -(1-2e^{-t})^{-1}.
} $$

For the remainder of this section, let  $D$ denote the Dirac operator
with
respect to this new metric and denote by ${\widetilde D}$ the Dirac
operator
associated to the old metric ${\tilde g} = e^{2log a_2} g.$  We
recall (see
\rLM\  ) that there is a local isometry $\phi$ between the spaces of
spinors
determined by the two conformal structures  such that
$${\widetilde D} s = e^{-3loga_2/2}D(e^{3log a_2/2}\phi(s)).$$
Hence, the map
$$\Phi: s\rightarrow e^{3loga_2/2}\phi(s)$$
takes harmonic spinors to harmonic spinors. This would induce an
isomorphism
between spaces of $L^2$  harmonic spinors if $\Phi$ preserved the
$L^2$
condition.
\vskip 0.1in
{\noindent \bf Proposition 3.3.1} {\it The map $\Phi$ is an
isomorphism between
spaces of $L^2$ harmonic spinors.}

\noindent
Proof.  Let $h\in L^2({\widetilde g})$.  Then
$$\infty > \int|h|^2dv_{\tilde{g}} \sim \int|h|^2e^{4t}dv_{g}.$$
Hence, $\int|e^{3t/2}\phi(h)|^2e^{t}dv_{g} < \infty,$ and $\Phi$ maps
$L^2$
harmonic spinors to $L^2$ harmonic spinors.  To obtain an
isomorphism, we must
also show that the inverse map also preserves the $L^2$ condition. It
suffices
to show that for
$H\in {\rm Ker} (D)$, $e^{t(1/2 + a)}H\in L^2(g)$ for some positive
$a$. We
shall prove this estimate below.  First we
 need a preliminary discussion of the holonomy of the index bundle,
$I^n$.

The fundamental group of $SO(3)/{\bf Z}_2$ is generated by the
inclusion
of the circle $K$ obtained by exponentiating the $Y_3$ vector.  This
is
the fiber of the fibration
\eqn\fiber{
\{L\}\times SO(3)/{\bf Z}_2\rightarrow \{L\}\times SO(3)/K{\bf Z}_2.}
According to \rMS, the holonomy about this circle fiber is given by
$e^{\pm i\pi n/2}$ when $L=\infty$. They show this by integrating
the curvature over the Atiyah-Hitchin cone \rAH.   Hence, for finite
$L$, the
holonomy differs
from this factor by $e^{i\pi \epsilon(L)}$, where $\epsilon$ is
doubly
exponentially decreasing. This can be seen by integrating the doubly
exponentially decreasing curvature over the subset of the cone with
$r > L$.
Note that the ${\bf Z}_2$ action prevents us from globally fixing the
sign
since $Y_3$ is only globally defined up to a factor of $\pm 1$.  We
will assume
$Y_3$ chosen so that the holonomy $\sim e^{- i\pi n/2 +
i\epsilon(L)}$.

It is more convenient to work with periodic frames than with
covariantly
constant ones.  Multiplying a $Y_3$-covariantly constant frame for
$I^n$ by
$e^{-i\psi(n/4 - \epsilon/2)}$ gives a connection form of ${in\over
4} Y_3^*$
modulo a doubly exponentially decreasing term.

Taking now a frame for the spinors determined by our (periodic) frame
$\{Y_0,Y_1,Y_2,Y_3\}$ and the periodic frame for $I^n$, we can
Fourier expand
sections in the $Y_3$ direction.  This can even be done globally,
although  the
Fourier coefficients then become sections of a ${\bf Z}_2$ quotient
of powers
of the Hopf bundle over $S^2$.

 Setting $T = Y_0$, we can write
$$D^2 = -T^2 - \nabla_{Y_1}^2 -\nabla_{Y_2}^2 - \nabla_{Y_3}^2 +
\nabla_{\nabla_{Y_3}Y_3},$$
modulo  rapidly decreasing terms. Then taking the $k^{th}$ Fourier
component
with respect to the circle action associated to $Y_3$, we have
$$
\eqalign{D^2 = & -T^2 + T - (e^t/2 - 1)^2(ik + in/4 -
1/2(1-[e^t/2-1]^{-2}/2 )
e_1e_2\cr
                             & - \frac{e^t}{4(e^t/2-1)^2}e_0e_3)^2
\cr
}$$
 plus a positive operator and rapidly decreasing terms.  This
operator is
conjugate to

$$\eqalign{&-T^2 + 1/4  - (e^t/2 - 1)^2(ik + in/4 -
1/2(1-[e^t/2-1]^{-2}/2 )
e_1e_2 \cr & - \frac{e^t}{4(e^t/2-1)^2}e_0e_3)^2\cr }$$
plus a positive operator and rapidly decreasing terms.  The smallest
eigenvalue of $- (ik + in/4- 1/2 e_1e_2)^2$ is $1/16$ when
$n\not\equiv 2$ (mod 4), since $k$ is an integer.
Hence, a standard maximum principle argument (or differential
inequality)
says that for
 $n\not\equiv 2$ (mod 4), any $L^2$
harmonic spinor is doubly exponentially decreasing and therefore maps
to an $L^2(\tilde{g})$ harmonic spinor under the previously described
map.
Moreover, this estimate implies that $D$ is Fredholm with no
continuous
spectrum in the case of restricted $ n$.  We are left to prove the
proposition
in the case $n\equiv 2$ (mod 4). In this case, $(ik + in/4- 1/2
e_1e_2)$
can have a kernel.  On this kernel, the $-\nabla_3^2$ term is
dominated by
$-(\frac{e^t}{4(e^t/2-1)^2}e_0e_3)^2 = 1/4$ modulo exponentially
decreasing terms. Hence, the decay of the component corresponding to
the kernel of $(ik + in/4- 1/2 e_1e_2)$ is, by a maximum principle
argument, of the order $O(e^{-t/2}).$ As the volume of the $SO(3)$
orbit at $t$
is
$O(e^{-t})$, $e^{t(1/2 + a)}H\in L^2(g)$ for $H\in {\rm Ker} (D)$.
 This implies that
$\Phi^{-1}$ takes $ L^2$ harmonic spinors to $L^2 $ harmonic spinors,
completing the proof of our proposition. ${\,\lower0.9pt\vbox{\hrule
\hbox{\vrule height 0.2 cm \hskip 0.2 cm \vrule height 0.2
cm}\hrule}\,}$

This proof shows that the essential spectrum of $D^2$ is bounded
away from zero.  Hence, $D$ is Fredholm.  A similar argument works to
show
when $n \not\equiv 2$ (mod 4) that $\widetilde D$ is Fredholm.

\subsec{Parametrices}

For the convenience of the reader, we gather in this section aspects
of the
construction of the approximate heat kernels.  The idea here is to
determine
an operator which is an appropriate approximation to $e^{-sD^2}$ for
use in
computing $\delta_D.$ We will define such an operator and find that
only a few
relevant components contribute to our calculation.

Using the functional calculus, we write
$$e^{-sD^2} = \frac{1}{2\pi i}\int_{\gamma} e^{-s\lambda}(D^2 -
\lambda)^{-1}d\lambda,$$
where $\gamma\subset C$ is a simple curve surrounding the spectrum
of $D^2$.  Thus, it suffices to approximate $(D^2 - \lambda)^{-1}.$

We construct the semilocal approximation to $(D^2 - \lambda)^{-1}$
inductively using a continuous Fourier transform in the base
variables
(in a coordinate neighborhood) and a discrete Fourier expansion in
the fibers.
More precisely, we consider  contractible open subsets of $RP^2$ (the
base of
the fibration) over which the fibration  is trivial.  As in section
3.3, we
Fourier expand the sections of the
twisted spinor bundles.  Fix a coordinate neighborhood $V$ on the
base space of
the fibration. On the $k^{th}$ Fourier component $D^2$ has the form
$$
\eqalign{ &-T^2 + T - \nabla_{Y_1}^2 -\nabla_{Y_2}^2 - (e^t/2 -
1)^2(ik + in/4
- 1/2(1-[e^t/2-1]^{-2}/2 ) e_1e_2\cr
 & - \frac{e^t}{4(e^t/2-1)^2}e_0e_3)^2 \cr }$$
plus rapidly decreasing terms.

Let
$$K^2 = (e^t/2 - 1)^2(ik + in/4 - 1/2(1-[e^t/2-1]^{-2}/2 )
e_1e_2)^2.$$
Let $\|v\|^2 = \sum g^{ij}v_{ij}.$ In a nice frame, let
$f = \sum_k f_k(x)e^{ik\psi}$ denote the Fourier expansion of a
section $f$,
and let $\hat f_k(v)$ denote the Fourier
transform of $f_k$.  Then the action of $D^2$  on $f$ is given
via the inverse Fourier transform by operating on $\hat f_k$ by
multiplication by $\|v\|^2 + K^2$ plus some lower order operator L,
which is
readily computed.  Hence, to invert $(D^2 - \lambda)$ approximately,
it suffices to invert
$(\|v\|^2 + K^2 -\lambda + L)$ approximately.  We construct such an
inverse
in the form $\sum_{l = 0}^N\frac{h_l}{(\|2\pi v\|^2 + K^2 -
\lambda)^{l+1}}.$
For the generic case, $K$ is very large and so high powers of
$(\|2\pi v\|^2 + K^2 - \lambda)^{-1}$ are rapidly decreasing.  Hence,
the
numerators
are constructed by an inductive process, so that
when acted on by $D^2 - \lambda$ (in the guise of $\|v\|^2 + K^2 +
L$)
we obtain $1$ plus a high power of $(\|2\pi v\|^2 + K^2 -
\lambda)^{-1}$.
The inductive construction is given as follows. Set
$$
h_0(x,x',\lambda ,v,k) = {\rm Identity}.
$$
Write
$$ \eqalign{   & (D^2 -\lambda )[(2\pi)^{-1}e^{ik\cdot
(\psi-\psi')}e^{i2\pi
(x-x')\cdot v}
(\|2\pi v\|^2 + K^2 -  \lambda )^{-l-1} h_l(x,x',\lambda ,v,k)  \cr =
\,
& (2\pi )^{-1}e^{ik\cdot (\psi-\psi')}e^{i2\pi (x-x')\cdot v} (\|2\pi
v\|^2 +
K^2 -
\lambda )^{-l}  h_l(x,x',\lambda ,v,k)  -  \cr & 2\nabla ((2\pi
)^{-1}
e^{ik\cdot (\psi-\psi')}  e^{i2\pi (x-x')\cdot v}) \cdot
\nabla \{(\|2\pi v\|^2  + K^2 - \lambda )^{-l-1}h_l(x,x',\lambda
,v,k)\} \cr &
+(2\pi )^{-1}  e^{ik\cdot (\psi-\psi')}e^{i2\pi (x-x')\cdot v}
\Delta_2
\{(\|2\pi v\|^2 + K^2 - \lambda )^{-l-1}  h_l(x,x',\lambda ,v,k)\}
\cr =\, &
(2\pi)^{-1}e^{ik(\psi-\psi')}e^{i2\pi(x-x')\cdot v}(\|2\pi v\|^2 +
K^2
-\lambda)^{-l}[h_l(x,x',\lambda,v,k) + R_l],
 }
$$
where $\Delta_2 = D^2 - K^2$.
\noindent
Set
$$
h_{l+1} = -(R_l)e^{ik\cdot (\psi-\psi')}e^{i2\pi (x-x')\cdot v}
(\|2\pi v\|^2 + K^2 - \lambda )^{l+1}.
$$
\medskip
\noindent
Then one obtains formally
$$
(D^2 -\lambda)\sum_k\sum_{l=0}^{N}\int_{{\bf R}^3}
(2\pi )^{-1}e^{ik\cdot (\psi-\psi')}e^{i2\pi (x-x')\cdot v}
\frac{h_l(x,x',\lambda,v,k)dv}{(\|2\pi v\|^2 + K^2 - \lambda )^{l+1}}
=
$$
$$
{\rm Identity} + \int_{{\bf R}^{3}}(R_N)dv,
$$
\noindent
and we take for some large $N$,
$$\sum_k\sum_{l=0}^{N}\int_{{\bf R}^3}
(2\pi )^{-1}e^{ik\cdot (\psi-\psi')}e^{i2\pi (x-x')\cdot v}
\frac{h_l(x,x',\lambda,v,k)dv}{(\|2\pi v\|^2 + K^2 - \lambda
)^{l+1}}$$
 for our semilocal approximation to $(D^2 - \lambda)^{-1}$ (pre- and
post-multiplied by cutoff functions in the usual way and summed over
a cover,
etc.).

Write
$$
Dh_l(x,x,\lambda,v,k) =
\sum_ {\sigma }
(\|2\pi v \|^2 + K^2 - \lambda )^{-\sigma }h_{l,\sigma }(x,v,k ),
$$
and
$$
h_{l,\sigma }(x,v,k) = \sum_{0\leq |A| + B\leq 2\sigma + 1,C\geq 0}
h_{l,\sigma,A,B,C}(x)v^A (e^tk)^Be^{-tC},
$$
\noindent
with $h_{l,\sigma,A,B,C}(x)$ bounded.

In the following section,  we shall study traces of the operator

$$ \eqalign{\int_0^{\infty}\int_{\gamma} e^{-s\lambda}\int_{{\bf
R}^3}
\frac{h_{l,\sigma,A,B,C}(x)v^A (e^tk)^Be^{-tC})dvd\lambda ds}{(\|2\pi
v\|^2 +
K^2 - \lambda )^{l+1+\sigma}}  & \leq }$$

$$ O((ke^{t})^{[B+|A|- 2l - 2\sigma - C + 1]})  \leq O((ke^{t})^{[2 -
2l - C]}.
 $$
Hence, we see without yet using the trace identities that only the
terms with
$l = 0$ or $1$ can contribute to the trace.  Let us now recall the
basic trace
lemma.

\vskip 0.1in
{\noindent \bf Lemma 3.4.1} {\it For $i_1,\cdots,i_j$ distinct and $j
> 0,$}
$$tr e_{i_1}\cdots e_{i_j} = 0.$$
So in computing the trace of $e_0\tau D h_l(x,x,\lambda,v,k)$ we get
a
nonzero contribution only from those terms which introduce enough
Clifford
factors to cancel $e_0\tau = -e_1e_2e_3$ -- and no more.  No term in
$Dh$
contributes a factor of $e_3$ without also contributing
an extra $e_0$ except the term
$$e_3(e^t/2 - 1)(ik + in/4 - 1/2(1-[e^t/2-1]^{-2}/2 ) e_1e_2)h$$
so this is the only term which can have nonzero trace.
We will compute the contribution of $Dh_0$ to this trace.  A similar
computation
shows that the contribution of $Dh_1$ is rapidly decreasing.

We compute
$$\eqalign{ & \frac{1}{2\pi
i}\int_{\gamma}e^{-t\lambda}\sum_k\int_{{\bf R}^3}
(2\pi)^{-1}e^{ik(\psi-\psi')}e^{i2\pi(x-x')\cdot
v}\frac{h_0(x,x',\lambda,v,k)dv d\lambda}{(\|2\pi v\|^2 + K^2 -
\lambda)} \cr
= & \frac{1}{2\pi i}\int_{\gamma}e^{-t\lambda}\sum_k\int_{{\bf R}^3}
(2\pi)^{-1}e^{ik(\psi-\psi')}e^{i2\pi(x-x')\cdot
v}\frac{dvd\lambda}{(\|2\pi
v\|^2 + K^2 - \lambda)}  \cr
= & \sum_k (4\pi
t)^{-3/2}e^{-|x-x'|^2}e^{-tK^2}(2\pi)^{-1}e^{ik(\psi-\psi')}.
\cr} $$
Recall we need to compute the trace of
$$\eqalign{& \sum_ke_0\tau e_3(e^t/2 - 1)(ik + in/4 -
1/2(1-[e^t/2-1]^{-2}/2 )
e_1e_2) (4\pi t)^{-3/2} \times \cr &
e^{-|x-x'|^2/4t}e^{-tK^2}(2\pi)^{-1}e^{ik(\psi-\psi')} \cr}$$
along $x = x'$, $\psi = \psi'$ where we recall that $ \tau$ is the
volume form
$ e_0 e_1 e_2 e_3$.  Integrating this trace over the $S^1$ fiber
reduces us to computing
\eqn\para{
\sum_k tr e_1e_2(e^t/2 - 1)(ik + in/4 - 1/2(1-[e^t/2-1]^{-2}/2 )
e_1e_2) (4\pi
t)^{-3/2}e^{-tK^2}.}

\subsec{Defect Computations}

We can now turn to the computation of  $ \delta_D$ defined in
\definition. Let
$e_i$ denote Clifford multiplication by
$Y_i$.  In our situation, we  have the following $L^2$ index theorem
as we have
previously discussed.
\vskip 0.1in
{\noindent \bf Proposition 3.5.1} {\it The index on $I^n$ is given by
the
expression}
$${\rm Ind} (D) = n^2/8 - 1/6 + lim_{L\rightarrow \infty}\int_{t =
L}\int_0^{\infty}tr e_0\tau De^{-sD^2}ds.$$
\medskip
The computation of $\delta_D$ is essentially the same as the index
computation in \rSS\ in the special case of a smooth divisor. Hence
we will restrict ourselves here to providing an outline of the
methods
involved.  The computation requires two steps: first we need to
construct an
explicit approximation $E_s$ to $e^{-sD^2}$ so that
$$lim_{L\rightarrow \infty}\int_{t = L}\int_0^{\infty}tr e_0\tau
D(e^{-sD^2}-E_s)ds = 0.$$
Using the results of the previous section, $E_s$ takes the form
(suppressing patching and partitions of unity etc.) :

$$\eqalign{ E_s(x,y) = & \, (2\pi
i)^{-1}\int_{\gamma}e^{-t\lambda}\sum_k\sum_{l=0}^N\int_{R^3}(2\pi)^{-
 1}
e^{ik(\psi-\psi')}e^{i 2\pi(x-x')\cdot  v}\times \cr & \,
\frac{h_l(x,x',\lambda,v,k)dv}
{(\|2\pi v\|^2 + K^2 -  \lambda)^{l+1}}.}$$
Second,  we need to compute
$$lim_{L\rightarrow \infty}\int_{t = L}\int_0^{\infty}tr e_0\tau
DE_sds.$$

For N large and $n \not \equiv 2 $ (mod 4), the presence of the
product of $-(e^t/2 - 1)^2$ and the square of the Fourier
coefficients in the
exponent can be used to show that $D(E_s - e^{-sD^2})$  has very
small trace
norm when multiplied by a cutoff function supported near $\infty$.

When $n \equiv 2$ (mod 4), this construction does not yield good
error estimates on the kernel of $(ik + in/4- 1/2 e_1e_2)$.  On this
subspace,
we must use a parametrix construction which is global on the entire
$SO(3)$ orbit.  This is similar to the construction of the parametrix
for singular m in \rSF\ -- see section 6 in particular.  There is a
natural
limiting Dirac
operator as $L\rightarrow \infty$ on this subspace, and we construct
our
heat operator (restricted to this subspace) as a perturbation of the
heat
operator associated to this
Dirac operator.  We need not compute this operator explicitly.  It
suffices to
note that it does not contribute any Clifford multiplication factors
to prevent
cancellation in the
super trace.  The necessary Clifford factors arise in the
perturbation
expansion, but with $O(e^{-L})$ coefficients. Hence this term
contributes
nothing to our trace.  The trace over the kernel of $(ik + in/4- 1/2
e_1e_2)$
 of the parametrix obtained by the semi-local construction also
contributes
zero.  Hence, we may use the semi-local construction for our entire
computation.

Set $U = (e^L/2 - 1)^{-1}$.
Returning to the computation of the traces
with the
semi-local parametrix, we see from \para\ that
$$\eqalign{  \int_{t = L}\int_0^{\infty}tr e_0\tau De^{-sD^2}ds = \,
&
 (4\pi)\sum_k\int_0^{\infty}tr e_1e_2U^{-1}(ik +
in/4 -
(1/2 - \cr &  U^2/4) e_1e_2)
e^{\frac{s}{U^2}(ik+in/4 - (1/2 - U^2/4)e_1e_2 )^2} \times \cr &
(4\pi
s)^{-3/2}
ds  + O(e^{-L}).   }$$
We simplify and apply the Poisson summation formula to write this as
$$ \eqalign{ &
 4\pi\int_0^{\infty}\sum_p\int_{\bf R}tr e_1e_2U^{-1}(ix +
in/4 - (1/2 - U^2/4)e_1e_2)  \times \cr &
e^{\frac{t}{U^2}(ix+in/4 - (1/2 - U^2/4)e_1e_2)^2}(4\pi t)^{-3/2}
e^{-2\pi
ipx}dxdt  + O(e^{-L}) \cr
=  \, &  4\pi\int_0^{\infty}\sum_p\int_{\bf R}tr e_1e_2U^{-1}ix
e^{\frac{-t}{U^2}x^2}(4\pi t)^{-3/2}  e^{-2\pi ipx} \times \cr &
e^{2\pi ip(n/4
+ i(1/2 - U^2/4)e_1e_2)}dxdt  + O(e^{-L})  \cr  = \, &
 4\pi tr\int_0^{\infty}\sum_p e^{2\pi ip(n/4 + i(1/2 - U^2/4)e_1e_2)}
e_1e_2 i
U^{-1}(-2\pi i)^{-1} \times \cr & \frac{\partial}{\partial p}
e^{-\pi^2p^2U^2/t}\pi^{1/2}Ut^{-1/2}
(4\pi t)^{-3/2}dt  + O(e^{-L}) \cr  = \, &
 tr\sum_{p\not = 0} e^{2\pi ip(n/4 + i(1/2 - U^2/4)e_1e_2)}
e_1e_2iU^{-1}(-2\pi
i)^{-1}\frac{\partial}{\partial p}
\frac{1}{2\pi^2p^2U}  + O(e^{-L}) \cr = \, &
 tr\sum_{p\not = 0} e^{2\pi ip(n/4 + i(1/2 - U^2/4)e_1e_2)}
  e_1e_2\frac{1}{2\pi^3 p^3 U^2}
   + O(e^{-L}). }$$
We now eliminate terms using the trace identities. Replacing
the factor $e^{\pi p(U^2/2e_1e_2)}$ by $e^{\pi p(U^2/2e_1e_2)} - 1$
does not change the trace, as $tr e_1e_2 1 = 0$. Moreover, we have
that
$$e^{\pi p(U^2/2e_1e_2)} - 1 = \pi p(U^2/2e_1e_2) + O(e^{-4L}).$$
Hence, the defect reduces to
$$
 tr\sum_{p\not = 0} e^{\pi ipn/2}
(-1)^p \pi pU^2 e_1e_2
  e_1e_2\frac{1}{4\pi^3 p^3 U^2}
  + O(e^{-L}),$$
which finally gives,
$$\delta_D =
 \sum_{p\not = 0} e^{\pi ipn/2}(-1)^{p+1} \frac{1}{\pi^2 p^2}.$$
We compute this for different values of $n$ (mod 4).
When $n = 0$,
$$\delta_D = \zeta(2)/\pi^2,$$
where $\zeta(s)$ denotes the
Riemann zeta function.
As is well known,
$\zeta(2) = \pi^2/6$. Hence, for $n = 0$ (mod 4)
$$\delta_D = 1/6.$$
\medskip
If $n = 1,$ (mod 4) we obtain
$$-\pi^{-2}\sum_{p\not = 0}p^{-2}e^{i\pi p3/2} = -\pi^{-2}\sum_{p
\not =
0}p^{-2}(-1)^p i^p = 1/2\sum_{p > 0}\pi^{-2}p^{-2}(-1)^{p+1} = 1/24
.$$
\medskip
If $n = 2,$ (mod 4) we obtain
$$-\pi^{-2}\sum_{p\not = 0}p^{-2}e^{i\pi p2} = -1/3 .$$
\medskip
If $n = 3,$ (mod 4) we obtain
$$-\pi^{-2}\sum_{p\not = 0}p^{-2}e^{i\pi p5/2} = -\pi^{-2}\sum_{p
\not =
0}p^{-2}i^p = - 1/2\sum_{p > 0}\pi^{-2}p^{-2}(-1)^p = 1/24 .$$

Before proceeding, let us summarize the results of this index
computation:

$$ {\rm dim Ker}(D_{I^n}) = n^2/8 - 1/6 +
\cases{\phantom{-}1/6,&$n\equiv
0\;{\rm mod}\; 4$\cr \phantom{-}1/24,&$n\equiv 1$\cr -1/3,&$n\equiv
2$\cr
\phantom{-}1/24, &$n\equiv 3$}$$

\subsec{Sections of ${\rm Ind}_2$ and Determining the Electric
Charges}

Until now we have been working on the space $M_2^0,$ although we are
actually
interested in sections of the index bundle ${\rm Ind}_2$ over
$(S^1\times M_2^0)/{I_3},$ where $I_3$ is a ${\bf Z}_2$ involution.
Since the
bundle is trivial over the ${\bf R}^3$ portion of the monopole moduli
space, we
can restrict our attention to the dependence of the bound state
wavefunction on
$S^1\times M_2^0.$ To obtain the full wavefunctions, we can simply
multiply by
a factor of $ e^{i \vec{p} \cdot\vec{x}},$ giving the particles
momentum.

To define sections of the quotient bundle ${\rm Ind}_2,$ one must
equivariantly
lift the ${\bf Z}_2$ action from the space $S^1\times M_2^0$ to the
total
space of the bundle $\widetilde{\rm Ind}_2$
over $S^1\times M_2^0.$  The $M_2^0$ piece of this bundle has been
discussed
in section 3.1.  Since $M_2^0$ is homotopic to $S^2,$ it is
convenient to
describe this piece of the bundle as a bundle over $S^2$ -- it is in
fact
the Hopf bundle (the total bundle over $M_2^0$
is then isomorphic to the pullback) \rMS.  The Hopf bundle $S^3
\rightarrow
S^2$
is the quotient map $\vec{z} \equiv (z_1,z_2) \mapsto
[z_1,z_2],$ where $\vert z_1\vert^2
+ \vert z_2\vert^2 = 1$ and $[z_1,z_2]$ is the point in ${\bf CP}^1
\cong S^2$
denoting the complex line through $  (z_1,z_2).$  The fiber is
clearly $U(1).$
Then by the homotopy,
we can think of the total space of the bundle ${\rm Ind}_2$
over $S^1\times M_2^0$ as a bundle $P$ over $S^1\times S^2,$ defined
as
follows \rMS:
\eqn\pbund{P = {\bf R}\times S^3 / \left\{  (t,\vec{z})\sim
(t+2\pi,-\vec{z})
\right\}.}
The bundle map is trivial on the first coordinate and the Hopf
map given above
on $\vec{z}.$  Note that on the base $t\sim t+2\pi,$ but we don't
have $(t,\vec{z})\sim (t+2\pi,\vec{z})$ on the total space.
Note, too, the nontrivial $S^1$ twist encoding the holonomy over the
$ S^1$
factor. The bundle over $M_2^0$ given by pulling back the Hopf bundle
in the
manner discussed above is just the bundle $ I$ of the previous
sections.
Now the relation \pbund\ is equivalently
\eqn\indbund{\widetilde{\rm Ind}_2 =  {\bf R}\times I/ \left\{
(t,e)\sim
(t+2\pi,-e) \right\},}
where $-e$ is $-1 \cdot e,$ with $-1$ acting along the fiber.

To get to ${\rm Ind}_2,$ we must quotient this bundle by ${\bf Z}_2.$
This procedure involves lifting $I_3$ to $\widetilde{I_3}$ acting
on the total space of $\widetilde{\rm Ind}_2.$  By ``lift,'' we mean
a
commutative
diagram
$$\matrix{P&\matrix{\widetilde{I_3}\cr \longrightarrow\cr
\phantom{X}}&P\cr
\downarrow&{}&\downarrow\cr
S^1\times S^2&\matrix{{I_3}\cr \longrightarrow\cr
\phantom{X}}&S^1\times
S^2.}$$
In describing the lift, we will for convenience work
with the bundle $P$ defined in \pbund\ and use the same notation
$\widetilde{I_3}$ and $I_3$ -- our
statements can be ``pulled
back'' to  $\widetilde{\rm Ind}_2.$  In terms of $S^2,$ the $I_3$ on
$M_2^0$
acts downstairs as the antipodal map on $S^2:$
$(z_1,z_2) \mapsto (-\overline{z}_2,
\overline{z}_1).$
Then $\widetilde{I_3}$ maps
$$\left(t,(z_1,z_2)\right) \rightarrow
\left( t + \pi , (-\overline{z}_2,\overline{z}_1 \right)).$$
One easily checks that $\widetilde{I_3}$ squares to the identity as a
result of the equivalence \pbund, and that the diagram is
commutative.
Now the quotient
$${\rm Ind}_2 \equiv \widetilde{\rm Ind}_2/ \widetilde{I_3}$$
makes sense as a bundle, with a well-defined bundle map
$\pi[(t,e)] = [(t,\pi(e))]$ because of the equivariance.

Recall how an automorphism acts on global sections:  if
$s: S^1\times M_2^0 \rightarrow \widetilde{\rm Ind}_2$ is
a global section, then
$$s \rightarrow \widetilde{I_3}^{-1}\circ s \circ I_3$$
under the action of $I_3.$ Let us use the same notation for the lift
of this
action to the bundle of spinors. Now this action commutes with the
Dirac
operator, so we can ask about the trace
of $I_3$ on the space of Dirac zero modes.  In fact, we will be able
to
glean this information in a simpler way which also reveals the
electrical
charges of the states we have found.

Note first that $\widetilde{I_3}$ defines a map $I_3^{(I)}$ which
acts only
on the $I$ factor of $\widetilde{\rm Ind}_2$ in \indbund.  Note that
$I_3^{(I)}$
squares to $-1,$ the map which is $-1$ on each fiber but trivial on
the base.
In fact, prior to this section we have only dealt with $I.$
Global sections of $\widetilde{\rm Ind}_2$ have the form
$\tilde{s}(t,m) = (t,s(t,m)) \in {\bf R} \times I.$ The $S^1$ piece
of the
Dirac equation dictates the $t-$dependence of $s$ to be
$s(t,m) = e^{iQt/2}s(m),$
where $s$ is a zero mode on $I$ and $Q$ is the electric charge.
Sections of
this form are clearly eigenstates of the electric charge operator $
- 2i
{\partial \over \partial t}.$
Note that multiplication by a complex phase along the fibers is
well-defined
on $I$ since it commutes with transition functions and is therefore
independent of trivialization (in a given trivialization, $s$ takes
the form
of a complex-valued function).

Now $\tilde{s}$ will descend to a global section on ${\rm Ind}_2$ if
$$\eqalign{\tilde{s}(t,m)
&= \widetilde{I_3} \circ \tilde{s} \circ I_3^{-1}(t,m) \cr
&= \widetilde{I_3} \circ \tilde{s}(t - \pi, I_3^{-1}m) \cr
&=  \widetilde{I_3}(t - \pi,e^{iQ(t-\pi)/2} I_3^{(I)}\circ
s \circ I_3^{-1}m)\cr
&= (t,e^{-iQ\pi/2}e^{iQt/2} I_3^{(I)}\circ s\circ I_3^{-1} m).}$$
We saw $(I_3^{(I)})^2 = -1$ on $I$ and since $I_3^{(I)}$ acts on the
space of
zero modes we can take $s$ to have definite eigenvalue equal to $r$
($r = \pm i$).  Putting $I_3^{(I)}\circ s\circ I_3^{-1}m = rs(m)$ and
reinstating the $t-$dependence yields
\eqn\eqcond{r(-i)^q = 1.}

In fact, this analysis holds for any odd power of $I.$  For even
powers,
the bundle of zero modes is untwisted over the $S^1$ piece and
therefore
$I_3^{(I)}$ squares to the identity.  Thus the eigenvalues $r$ take
the
form $r = \pm 1.$  The same condition \eqcond\ applies.
Thus:
$$\eqalign{n \;\;{\rm odd}  \,\,  \Rightarrow
&\, r = +i \;\;\hbox{states have charge}\;\; 1 \;\hbox{mod}\; 4 \cr
&\, r = -i \phantom{\hbox{states have}}\hbox{"}\phantom{
charge}\,\;\; 3
\;\hbox{mod}\; 4 \cr
n \;\;{\rm even}\Rightarrow
& \, r = +1  \;\;\hbox{states have charge}\;\; 0
\;\hbox{mod}\; 4\cr
& \, r = -1 \phantom{\hbox{states have}}\hbox{"}\phantom{
charge}\,\;\; 2
\;\hbox{mod}\; 4.}$$
In other words, every zero mode of $I$ yields a physical monopole
solution with
electric charge dependent upon its $I_3^{(I)}$ eigenvalue.

How do we count the number of solutions with a given charge?  We take
the
trace of $I_3^{(I)}$ on the space of zero sections.  This is done for
$n=4$
(the fourth power of $I$) below.  For $n=3$ we remark that there is
only one
eigenvector, with eigenvalue $\pm i,$ which must be the complex
conjugate
of the $n=-3$ eigenvector.  These states are charge conjugates (note
that $1$ and $-1$ have different odd values mod 4).  This argument
tells
us nothing for the even states, but for $n < 2$ there are no
solutions
anyway.

Let $L(I_3)$ denote the trace of $I_3^{(I)}$ restricted to the kernel
of $D$ on $I^4.$  In order to compute $L(I_3)$ we need a noncompact
variant of
the
Atiyah-Segal-Singer equivariant index theorem. We may argue exactly
as we
did in section 3.2 to obtain the following proposition. (See \rSS.)

\vskip 0.1in
{\noindent \bf Proposition 3.6.1}
$$ \eqalign{ L(I_3) = &
lim_{L\rightarrow\infty}(lim_{s\rightarrow 0}\int_{r<L}\tr\tau
I_3^{(I)}e^{-sD^2}dx
+ \int_{0}^{\infty}\int_{r = L} tre_0  \times \cr  & \tau
I_3^{(I)}De^{-sD^2}d\sigma ds).}$$
\medskip

The small $s$ limit can be evaluated exactly as in the compact case.
It
localizes to a computation in a neighborhood of the fixed point set
of $I_3$,
but $I_3$ has no fixed points.  Hence (see, for example, Section 6.3
of
\rBGV),

$$lim_{L\rightarrow\infty}(lim_{s\rightarrow 0}\int_{r<L}\tr\tau
I_3^{(I)}e^{-sD^2}dx) =0.$$
We are left to compute

$$lim_{L\rightarrow\infty}\int_{0}^{\infty}\int_{r = L} tre_0\tau
I_3^{(I)}De^{-sD^2}d\sigma ds.$$

We compute this defect term
using the same semilocal parametrix we used in section 3.5.
The action of $I_3^{(I)}$ changes the computation in two ways.
First, it
acts by $\pm e_1e_2$ on the spinors. The sign is determined by the
choice of
the lift of the action of $I_3$ to the principal spin bundle.
Secondly,
$I_3^{(I)}$ introduces introduces a rotation of $\pi$ in the fiber of
the
fibration \fiber.  This enters the computation by having us evaluate
the
parametrix not on the diagonal, but along the diagonal on the
base of the fibration and  on $(\psi,\psi + \pi)$ in the fiber. For
the $k^{th}$ Fourier component, this is
$(-1)^k$ times what one obtains by evaluating along the diagonal.
So $I_3^{(I)}$ changes our computation by introducing fators of
$e_1e_2(-1)^k$.  Once again
trace identities make all terms vanish except

$$ \eqalign{& -\int_{RP^2}\sum_k\int_0^{\infty}tr e_1e_2e_3
e_1e_2(-1)^ke_3(e^L/2 - 1)  (i[k +  1] - (1/2 - U^2/4)e_1e_2) \times
\cr &
e^{\frac{s}{2u}(i[k+1] -  (1/2 - U^2/4)e_1e_2)^2}(4\pi
s)^{-3/2}
ds  }$$
$$ \eqalign{= & -(2\pi)\sum_k\int_0^{\infty}tr (-1)^k(e^L/2 - 1)(ik
 - (1/2 - U^2/4)e_1e_2) e^{\frac{s}{2u}(ik -  (1/2 - U^2/4)e_1e_2)^2}
\times
\cr & (4\pi
s)^{-3/2} ds.}$$
This expression is skew under the involution that interchanges the
eigenspaces
of $e_1e_2$ and sends $k\rightarrow -k$.  Hence the trace is zero.

We can therefore conclude that for $ n=4$
\eqn\trace{L(I_3) = 0,}
and for $n= -4$ we also find $L(I_3) = 0.$
So there are an equal number of charge $0\;{\rm mod}\; 4$ and
$2\;{\rm mod}\;
4$ two-monopole states.

\newsec{The BPS Spectrum }

Let us briefly review the predictions of Seiberg and Witten and
compare our
results with their predictions. Of particular interest to us are the
global
symmetry properties of the theory. It will be an important check that
the BPS
states we find appear in predicted representations of the global
symmetry
group. Let us first exhibit the flavor symmetry discussed in \rSW\
explicitly.
For the purpose of determining the global symmetry, the Lagrangian
contains a
kinetic term for the hypermultiplets given by the bilinear form

$$ M^\dagger M + \Mt^{\dagger} \Mt,$$
where $N_f$ flavor indices as well as the gauge group indices are
suppressed.
The chiral superfields $ M$ and $ \Mt$ are in conjugate
representations of the
gauge group. However, since the gauge group is $ SU(2)$, the
fundamental and
anti-fundamental are isomorphic, so a symmetry   mixing $M$ and $\Mt$
is
permitted. The other term to be preserved in the Lagrangian  is the
coupling to
the Higgs field $\Mt^T \Phi M.$  If we define the $2N_f$-dimensional
complex
vector $$V \equiv \pmatrix{M + \Mt \cr i(M-\Mt)},$$ then a symmetry
$V
\rightarrow AV$ must preserve $V^\dagger V$ and $V^T V$  i.e.
$A\in U(2N_f)$ and $A\in O(2N_f;{\bf C}).$  So $A^\dagger = A^{-1} =
A^T,$ and
thus $A = A^*;$ therefore $A$ is in $O(2N_f)$ with real coefficients.
At the
quantum level, the parity in $ O(2N_f)$ either reverses the sign of
the
electric charges ($N_f $ odd), or is broken ($ N_f$ even). The
relevant global
symmetry group for states of a given charge is therefore $ SO(2N_f).$

As we discussed in section 2, the low-energy dynamics of monopoles
and dyons is
described by a supersymmetric sigma model.  The Hilbert space
decomposes into
representations of $ SO(2N_f).$ For magnetic charge $ k=1$, the
bundle of zero
modes is one-dimensional, and so the zero mode anti-commutation
relations

$$ \left\{ \psi^i , \psi^j \right\} = \delta^{ij}  \,\,\,\,\,\, i=1,
\ldots ,
2N_f $$
lead to spinorial representations of $ SO(2N_f)$ as noted in \rSW.
For magnetic
charge $ k=2$, the bundle is two-dimensional, and so the algebra that
we must
represent in terms of flavor properties is no longer a Clifford
algebra but
\eqn\algebra{ \left\{ \psi^i , \psi^{j \dagger} \right\} =
\delta^{ij},  }
where the $ \psi^i$ are now complex. This is just the usual
annihilation and
creation operator algebra. BPS states with magnetic charge two
therefore fall
into representations of $ SO(2N_f)$ rather than its universal cover.

For  $ N_f=0,1,2$, we found no BPS bound states with magnetic charge
two, and
the singularity structure proposed by Seiberg and Witten required no
bound
states. In the case $ N_f=0$, the semi-classical  BPS spectrum is
completely
determined since there are no bound states for any magnetic charge $
k>1$. Such
states would correspond to  normalizable anti-holomorphic forms on a
non-compact Calabi-Yau manifold, and there are no such forms.

For $ N_f=3$, we found a single bound state for each charge $ (2, n)$
where $
n$ is any odd integer. These states are singlets under the $ SO(6)$
flavor
symmetry since they correspond to the Fock vacuum  for the algebra
\algebra\
and its complex conjugate.  In this case, the singularity structure
of the
moduli space can be analyzed by looking at the limit of three equal
very
massive quarks.  Since the mass is large,  there is a singularity in
the
semi-classical region of the moduli space where the scalar vacuum
expectation
value is large and the quarks (in the ${\bf 3}$ of the $SU(3)$ flavor
symmetry)
become massless.  In the strong coupling region one can integrate out
the
massive quarks semiclassically and relate the singularities to the
two known
singularities -- a monopole with $(n_m,n_e) = (1,0)$ and a dyon
$(1,1)$ both in
the ${\bf 1}$ of $SU(3)$ -- of the $N_f = 0$ quantum moduli space.
In this
way, the nonabelian global charges of the massless states at the
singularities
are calculated.  These charges cannot change.  One can then let the
masses of
the quarks go to zero and determine the representation theory under
the full
$\hbox{Spin}(6) = SU(4)$ symmetry.  This requires the ${\bf 3}$ and
${\bf 1}$
to combine into a ${\bf 4}$ singularity, while the other singularity
(now a
${\bf 1}$ under $SU(4)$) remains separate.  For $SU(4)$ the center
${\bf Z}_4$
is determined by general consistency conditions to act as $e^{i\pi
(n_m+2n_e)/2},$ yielding the condition for the ${\bf 4}$ that
$n_m+2n_e = 1,$
with minimal solution $(n_m,n_e) = (1,0).$ This just the standard
$(1,0)$ BPS
monopole.  The singlet state is trivial under the center and thus has
minimal
solution $(n_m,n_e) = (2,1).$  As explained in \rSW, this state
should be
continuously connected to a BPS state which exists semi-classically,
and we
have found such a state.

The case of most interest is $ N_f=4.$ This theory is conjectured by
\rSW\ to
be self-dual under $SL(2,{\bf Z}).$  However, the situation is
somewhat
different from the N=4 \YM\  theory which is also conjectured to be
self-dual.
For N=2, $N_f = 4,$ the $SL(2,{\bf Z})$ is believed to act on  the
representation spaces of the ${\rm Spin}(8)$ global symmetry as well.
The
elementary hypermultiplets transform in the vector of ${\rm Spin}(8)$
(i.e. the
fundamental of $SO(8)$), while the $(1,0)$ and $(1,1)$ monopole
states
transform in opposite chirality spinor representations.  Although
these
representations are not isomorphic, they are all eight dimensional
and permuted
by the ${\bf S}_3$ which acts as outer automorphisms of ${\rm
Spin}(8)$ (an
inner automorphism would give an isomorphic representation).  The
trivial
representation is trivial under this ${\bf S}_3$ as well.  We will
not further
motivate this prediction, but will rather simply state the action of
symmetry.
The action of $SL(2,{\bf Z})$ on a state is to transform the monopole
numbers
in the usual  way -- $(n_m,n_e)\rightarrow (an_m+bn_e,cn_m+dn_e)$
under
$\pmatrix{a&b\cr c&d}\in SL(2,{\bf Z})$ -- and the representation is
transformed under $\rho \in {\bf S}_3$ as above, where $\rho$ is the
matrix
$\pmatrix{a&b\cr c&d}\in SL(2,{\bf Z})$ with entries mod 2 (the group
of unit
determinant matrices mod 2 is easily seen to be isomorphic to ${\bf
S}_3$)
acting by left multiplication on the representations $(0,0) =$
trivial $o$,
$(0,1) =$ vector $v$, $(1,0) =$ spinor $s,$ $(1,1) =$ spinor $c,$
where all
numbers are defined mod 2.  Simply stated:  the representations are
determined
by $(n_m, n_e )$ mod 2.

The BPS spectrum contains a stable elementary electron with charge
$(0,1)$.
The self-duality conjecture then implies the existence of bound
states for all
charges $ (p,q)$ where $ p$ and $ q$ are relatively prime integers.
Specifically, bound states with charge $ (2,q)$ with $ q$ odd must
exist and
appear in the vector representation of  ${\rm Spin}(8)$. We indeed
found such
states from the bound state solution corresponding to the excitation
of a
single zero mode $ \gamma^{i \dagger}|-4>$ on the vacuum with $ U(1)$
charge $
-4$, and its complex conjugate. The allowed electric charges are $
4q+1$ for
one state, and $ 4q+3$ for the other.

The BPS spectrum also contains neutrally stable heavy gauge bosons of
charge $
(0,2)$.  If such states exist as discrete states in the theory then
we should
expect to see their partners under $SL(2,{\bf Z})$ with charges $
(2p,2q)$ as
discussed in \rSW. These states must all be singlets under ${\rm
Spin}(8)$.
Further, the heavy gauge bosons are part of BPS multiplets with spins
$ \leq 1$
unlike the electrons which are part of  BPS multiplets with spins $
\leq
{1\over 2}$. From our computations, we have shown that  four bosonic
bound
state solutions exist that are singlets under   ${\rm Spin}(8).$ We
found in
\trace\ that the electric charges are $ 4q$ for two of  the solutions
and $
4q+2$ for the remaining two.  The existence of such states certainly
implies
that if the theory is self-dual then the heavy gauge bosons must
exist as bound
states at the threshold of decay into electrons. Two solutions are
also
required if one is to construct a BPS multiplet containing a vector
particle.
Our findings are certainly in accord with the proposed duality.

To further support the supposition that the two solutions at $ n=\pm
4$ with a
given electric charge are members of the same BPS multiplet, we can
examine the
difference in fermion number between the different bound state
solutions.  We
begin by noting that  in the models under consideration, the fermion
number is
always integral. The fermion number of a bound state solution comes
from two
sources \rJR: the first contribution is from fermions in the
effective action
for pure N=2 \YM.   Since the moduli space is a product, the action
can always
be written as  the sum of two terms: $ S_{eff}(k=1)$ and an
interacting piece $
S_{int}$. Quantizing the fermions from the first term gives us the
usual
four-dimensional BPS multiplet when acting on a spin zero vacuum.
Our
interest resides with the remaining fermions from $ S_{int}$, and the
difference in the fermion number between the $ n=3$ and $ n=4$ bound
state
solutions. Viewing these fermions as spinors, and noting that the
index has the
same sign for $ n=3$ and $ n=4$, we can conclude that the bound state
solutions
at $ n=3$ and $ n=4$ have the same chirality. Therefore, the
difference in
fermion number from this source is zero mod 2. The other
contribution, from the
matter fermions, clearly produces a difference in fermion number.
Therefore,
we can conclude that the bound states at $ n=4$ differ in fermion
number from
the bound state at $ n=3$ by one mod 2.   Spin-statistics implies
that the
bound state at $ n=3$ is bosonic, and so the two states at $ n=4$ are
fermionic. A BPS multiplet built on such a vacuum includes a vector
particle as
expected.

\newsec{Conclusions}

We have studied the question of whether bound states of monopoles and
dyons
with magnetic charge two exist in supersymmetric \YM\ coupled to
matter. This
problem was solved by computing the number of  $L^2 $ solutions of
the Dirac
equation for bundle-valued spinors over the two-monopole moduli
space. For $
N_f  <3$, no bound states exist, while for $ N_f=3$, there is a
single bound
state for every odd value of the electric charge,  which is a singlet
under the
$ SU(4)$ symmetry group. For $ N_f=4$, there is a bound state for
each odd
value of the electric charge. These bound states are in the vector
representation of the ${\rm Spin}(8)$ flavor symmetry. There are also
two bound
states for each even value of the electric charge, which are singlets
under the
flavor group.  Our findings provide dynamical evidence for the moduli
space
structure proposed by Seiberg and Witten -- specifically, for  the $
N_f=4$
conjectured self-duality.   To show the BPS spectrum of the $ N_f=4$
theory is
truly  $ SL(2,\bf{Z})$ invariant, similiar calculations are needed
for higher
magnetic charge.  The main obstacle is the limited information about
the metric
for the higher charge monopole moduli spaces. Recently, the
asymptotic metric
on the $ k$-monopole moduli space for the region where all $ k$
monopoles are
far apart has been described \rGMtwo.   However, an understanding of
the metric
at the boundaries of codimension one would be desirable to extend
computations
of this type.

\bigbreak\bigskip\bigskip\centerline{{\bf Acknowledgements}}\nobreak

We would like to thank J. Harvey, N. Hitchin, B.J. Schroers, I.
Singer, C. Vafa
and S.-T. Yau  for discussions. The work of S. S. was supported in
part by a
Hertz Fellowship, NSF grants PHY-92-18167, PHY-89-57162 and a Packard
Fellowship; that of M. S. by NSF Grant DMS 9505040; and that of E. Z.
by grant
DE-FG02-88ER-25065.

\listrefs

\end